\documentstyle[aps,12pt]{revtex}

\begin{document}

\preprint{EFUAZ FT-97-49}

\title{The Second-Order Equation from the $(1/2,0)\oplus
(0,1/2)$ Representation of the Poincar\'e
Group\thanks{Submitted to ``Acta Physica Polonica B"}}

\author{{\bf Valeri V. Dvoeglazov}}

\address{Escuela de F\'{\i}sica, Universidad Aut\'onoma de Zacatecas\\
Apartado Postal C-580, Zacatecas 98068 Zac., M\'exico\\
Internet address: VALERI@CANTERA.REDUAZ.MX\\
URL: http://cantera.reduaz.mx/\~~valeri/valeri.htm
}

\date{October 17, 1997}

\maketitle

\begin{abstract}
On the basis of the first principles we derive the Barut-Wilson-Fushchich
second-order equation in the $(1/2,0)\oplus (0,1/2)$ representation.
Then we discuss the  possibility of the description of various mass
and spin states in such a framework.
\end{abstract}

\bigskip

%\newpage

Few decades ago Prof. A. Barut and collaborators proposed to use
the four-dimensional representation of the $O(4,2)$ group in order to
solve the problem of the lepton mass splitting~\cite{Bar0,Wilson,Bar}.
Similar research has been produced by Prof. V. Fushchich and
collaborators~\cite{Fush}.

The most general conserved current that is linear in the generators of
the four-dimensional representation of the group $O(4,2)$ was given
as~\cite{Bar0,Wilson}
\begin{equation}
j_\mu = \alpha_1 \gamma_\mu + \alpha_2 P_\mu +\alpha_3 \sigma_{\mu\nu}
q^\nu \,\,,
\end{equation}
where $P_\mu =p_{1\mu} +p_{2\mu}$ is the total momentum, $q_\mu = p_{1\mu}
- p_{2\mu}$ is the momentum transfer and $\alpha_i$ are real
constant coefficients which may depend on the internal degrees of freedom
of leptons.   The Lagrangian formalism and the secondary quantization
scheme have been given in ref.~\cite{Wilson}.  A. Barut derived the mass
spectrum of leptons~\cite{Bar} after taking into account the additional
postulate that one has to fix the value of the anomalous magnetic moment
of the particle by its {\it classical} value $g=2(2\alpha /3)$, $\alpha$
is the fine structure constant.

Recently, we have come across the flexibility of the introduction of
similar constructs in the $(j,0)\oplus (0,j)$ representations from very
different standpoints~\cite{Dvo,Dvo1}.  Starting from the explicit form
of the spinors which are eigenespinors of the helicity
operator~\cite{Var} one can reveal {\it two non-equivalent} relations
between zero-momentum 2-spinors. The first form of the Ryder-Burgard
relation\footnote{This name was introduced by D.
V. Ahluwalia  when considering the $(1,0)\oplus (0,1)$
representation~\cite{Dva}. If one uses $\phi_{_R} (\overcirc{p}^\mu) =
\pm \phi_{_L} (\overcirc{p}^\mu)$, cf. also~\cite{Faust,Ryder}, after
application Wigner rules for the boosts of the 2-spinors to the momentum
$p^\mu$, one immediately arrives at the Bargmann-Wightman-Wigner type
quantum field theory, ref.~\cite{BWW} (cf. also~\cite{Gel}), in this
representation.} was given in ref.~\cite{AV} ($h$ is a quantum number
answering for the helicity; in general, see the cited papers for  the
notation) :
\begin{equation}
[\phi_{_L}^h (\overcirc{p}^\mu)]^\ast =
e^{-2i\vartheta_h} \Xi_{[1/2]} \phi_{_L}^h (\overcirc{p}^\mu)\,\,;
\end{equation}
the second form, in ref.~\cite{DVON}:
\begin{equation}
[\phi_{_L}^h (\overcirc{p}^\mu)]^\ast = (-1)^{{1\over 2} -h} e^{-i
(\vartheta_1 +\vartheta_2)}  \Theta_{[1/2]} \phi_{_L}^{-h}
(\overcirc{p}^\mu)\,\,.
\end{equation}
The matrices are defined in the $(1/2,0)\oplus (0,1/2)$ representation
as follows:
\begin{eqnarray}
\Theta_{[1/2]} = \pmatrix{0&-1\cr
1&0\cr}\,\,,\quad
\Xi_{[1/2]} = \pmatrix{e^{i\phi}&0\cr
0&e^{-i\phi}\cr}\,\, ,
\end{eqnarray}
here $\phi$ is the azimuthal angle related with ${\bf p}
\rightarrow {\bf 0}$.

In this paper we advocate the generalized Ryder-Burgard relation:
\begin{equation}
\phi_{_L}^h (\overcirc{p}^\mu) =
a (-1)^{{1\over 2} - h} e^{i(\vartheta_1 +\vartheta_2)} \Theta_{[1/2]}
[\phi_{_L}^{-h} (\overcirc{p}^\mu)]^\ast + b e^{2i\vartheta_h}
\Xi^{-1}_{[1/2]} [\phi_{_L}^h (\overcirc{p}^\mu)]^\ast
\quad,
\end{equation}
with the {\it real} constant $a$ and $b$ being arbitrary at this stage.
The relevant relations for $\phi_{_R}$ are obtained after taking into
account that
\begin{mathletters}
\begin{eqnarray}
\phi_{_L}^\uparrow (p^\mu) &=& - \Theta_{[1/2]} [\phi_{_R}^{\downarrow}
 (p^\mu)]^\ast \quad,\quad \phi_{_L}^\downarrow (p^\mu) = + \Theta_{[1/2]}
[\phi_{_R}^{\uparrow}(p^\mu)]^\ast \quad,\quad\label{1a}\\
\phi_{_R}^\uparrow (p^\mu) &=& -
\Theta_{[1/2]} [\phi_{_L}^{\downarrow} (p^\mu)]^\ast \quad,\quad
\phi_{_R}^\downarrow (p^\mu) = + \Theta_{[1/2]}
[\phi_{_L}^{\uparrow} (p^\mu)]^\ast \label{1b}\,\,,
\end{eqnarray}
\end{mathletters}
which are easily derived by means of the consideration of the explicit
form of 4-spinors, e.~g., ref.~\cite{Ryder}.\footnote{In fact, we have
certain room in the definitions of the right spinors due to the
arbitrariness of the phase factors at this stage. But, if one chosen the
spinorial basis as in~\cite{Ryder} one can find the relevant spinors in
any frame after the application of the Wigner rules \begin{equation}
\phi_{_{R,L}} (p^\mu) = \exp (\pm {\bbox \sigma}\cdot {\bbox \varphi} /2)
\phi_{_{R,L}} (\overcirc{p}^\mu)\quad.  \label{2}
\end{equation} for right
(left) spinors); $\cosh (\varphi) = E/m$, $\sinh (\varphi) = \vert {\bf p}
\vert /m$, with $\widehat{\varphi} ={\bf p}/\vert {\bf p}\vert$ is the
unit vector.  In the subsequent papers we shall consider different choices
of the phase factors between left- and right- spinors.} Next we apply the
procedure outlined in ref.~\cite[footnote \# 1]{AV}.  Namely,
\begin{eqnarray}
\phi_{_L}^h (p^\mu) &=& \Lambda_{_L} (p^\mu \leftarrow
\overcirc{p}^\mu)) \phi_{_L}^h (\overcirc{p}^\mu) =
\Lambda_{_L} (p^\mu \leftarrow \overcirc{p}^\mu ) \left \{ a (-1)^{{1\over
2}-h} e^{i(\vartheta_1 + \vartheta_2)} \Theta_{[1/2]} [\phi_{_L} ^{-h}
(\overcirc{p}^\mu) ]^\ast +\right. \nonumber\\
&+& \left. b e^{2i\vartheta_h} \Xi^{-1}_{[1/2]}
[\phi_{_L}^h (\overcirc{p}^\mu)]^\ast \right \} =
-a e^{i(\vartheta_1 +\vartheta_2)} \Lambda_{_L} (p^\mu \leftarrow
\overcirc{p}^\mu ) \Lambda_{_R}^{-1} (p^\mu \leftarrow \overcirc{p}^\mu)
\phi_{_R}^h (p^\mu) +\nonumber\\
&+& b e^{2i\vartheta_h} (-1)^{{1\over 2}
+h} \Theta_{[1/2]} \Xi_{[1/2]} \phi_{_R}^{-h} (p^\mu) \,\,.\label{3}
\end{eqnarray}
As a consequence of Eqs.  (\ref{1a},\ref{1b},\ref{2},\ref{3})
after the choice of the phase factors (e.~g., $\vartheta_1 =0$,
$\vartheta_2 = \pi$) one has
\begin{eqnarray}
\phi_{_L}^h (p^\mu) &=& a \frac{p_0 -\bbox{\sigma}\cdot {\bf
p}}{m}\phi_{_R}^h (p^\mu) +b (-1)^{{1\over 2}+h}\Theta_{[1/2]} \Xi_{[1/2]}
\phi_{_R}^{-h} (p^\mu)\,\, ,\\
\phi_{_R}^h (p^\mu) &=& a \frac{p_0 +\bbox{\sigma}\cdot {\bf
p}}{m}\phi_{_L}^h (p^\mu) +b (-1)^{{1\over 2}+h}\Theta_{[1/2]} \Xi_{[1/2]}
\phi_{_L}^{-h} (p^\mu)\,\, .
\end{eqnarray}
Thus, the momentum-space Dirac equation is generalized:
\begin{equation}
(a{\widehat p \over m} -\openone) u_h (p^\mu) +i b (-1)^{{1\over 2}-h}
\gamma^5 {\cal C} u_{-h}^\ast (p^\mu) =0\,\,,
\end{equation}
where
\begin{eqnarray}
{\cal C} = \pmatrix{0&i\Theta_{[1/2]}\cr -i\Theta_{[1/2]}&0\cr}\,\, ,
\end{eqnarray}
the $4\times 4$ matrix, which enters in the definition of the charge
conjugation operation. The counterpart in the coordinate space is
\begin{equation}
\left [a \,{i\gamma^\mu \partial_\mu \over m}
+b\, {\cal C} {\cal K} - \openone\right ] \Psi (x^\mu) = 0\,\, ,\label{de}
\end{equation}
provided that in the operator formulation the creation (annihilation)
operators are connected by
\begin{equation}
b_\downarrow (p^\mu) = - i a_\uparrow (p^\mu)\quad,\quad
b_\uparrow (p^\mu) = + i a_\downarrow (p^\mu)\,\, ,
\end{equation}
We denote ${\cal K}$ to be the operation of the complex conjugation  but
one should imply that it acts as the Hermitian conjugation on
the creation (annihilation) operators in the $q-$ numbers theories.

On the other hand we are aware about the possibility of dividing the Dirac
function into the real and imaginary part~\cite{Maj}. The transformation
from the Weyl representation of the $\gamma^\mu$ matrices is made by means
of the unitary $4\times 4$ matrix
\begin{equation}
U ={1\over 2}\pmatrix{\openone -i\Theta_{[1/2]} & \openone
+i\Theta_{[1/2]}\cr -\openone -i\Theta_{[1/2]} & \openone
-i\Theta_{[1/2]}\cr}\quad,\quad U^\dagger = {1\over 2}\pmatrix{\openone
-i\Theta_{[1/2]} & -\openone -i\Theta_{[1/2]}\cr \openone +
i\Theta_{[1/2]} & \openone -i\Theta_{[1/2]}\cr}\,\,.
\end{equation}
The similar transformations can also be applied to the Weinberg (or
Weinberg-Hammer-Tucker) spin-1 equation in order to divide the spin-1
function into the real and imaginary parts~\cite{Dvo2}.  Presumably, this
is the general property of all fundamental wave equations.

Let us try to write the generalized coordinate-space Dirac equation
(\ref{de}) in the Majorana representation:
\begin{equation}
\left [ a \,{i\gamma^\mu \partial_\mu \over m} -b\,{\cal K} -\openone
\right ] \Psi (x^\mu) =0\,\,.
\end{equation}
We used that $U{\cal C}
{\cal K} U^{-1} = -{\cal K}$. After Majorana we note that $\gamma^\mu$
matrices become to be {\it pure imaginary} and the Dirac function can be
divided into $\Psi \equiv \Psi_1 +i\Psi_2$.  Hence, we have a set of two
{\it real} equations:
\begin{mathletters} \begin{eqnarray} \left [ a
{i\gamma^\mu \partial_\mu \over m} -b -\openone \right ] \Psi_1 (x^\mu)
&=&0\,\,,\\
\left [ a {i\gamma^\mu \partial_\mu \over m} +b -\openone
\right ] \Psi_2 (x^\mu) &=&0\,\,.
\end{eqnarray} \end{mathletters}
Adding and subtracting the obtained equations we arrived at the set
($\phi =\Psi_1 +\Psi_2$ and $\chi=\Psi_1 -\Psi_2$)
\begin{mathletters}
\begin{eqnarray}
\left [ a {i\gamma^\mu \partial_\mu \over m} -\openone \right ] \phi - b
\,\chi &=& 0 \,\,,\\
\left [ a {i\gamma^\mu \partial_\mu \over m} -\openone \right ]\chi  - b
\,\phi &=& 0\,\,,
\end{eqnarray}
\end{mathletters}
which after multiplication by $b\neq 0$ yield
the same second-order equations for $\phi$ and $\chi$:
\begin{equation}
\left [2a {i\gamma^\mu \partial_\mu \over m}  + a^2 {\partial^\mu
\partial_\mu \over m^2} +b^2 -\openone \right ] \cases{\phi (x^\mu) &\cr
\chi (x^\mu) &\cr} = 0\quad.  \label{Barut}
\end{equation}
With the identification ${a\over 2m} \rightarrow \alpha_2$ and ${1-b^2
\over 2a} m \rightarrow \kappa$ we obtain the equation (6) of ref.~[3a].
This is the equation which Barut used to obtain the mass difference
between a muon and an electron.

On using the similar technique and after the consideration of
different {\it chirality} sub-spaces as independent ones one can come to
the Fushchich equation~[4c]. It has the interaction with the 4-vector
potential which is the same as the $j=0$ Sakata-Taketani particle.
This problem is closely related with the recent discussion of the Dirac
equation with {\it two} mass parameters~\cite{Rasp}.
It seems to me that the theoretical difference (comparing with
the first-order Dirac equation), which we obtain after switched-on
interactions, is caused by the induced asymmetry between evolutions
toward and backward in the time (or, which might be equivalently,
between  different chirality/helicity sub-spaces).

In conclusion, we can now assert that the Barut guess was not
some puzzled coincidence. In fact, the second-order equation can be
derived on the basis of the minimal number of postulates (Lorentz
invariance and relations between 2-spinors in the zero-momentum frame) and
it is the natural consequence of the general structure of the $(j,0)\oplus
(0,j)$ representation.

{\it Acknowledgments.} The work was motivated by the papers
of Prof. D. V. Ahluwalia, by very useful frank discussions
and phone  conversations with Prof. A. F. Pashkov during last 15 years,
by Prof. A. Raspini who kindly sent me his papers and by the critical
report of anonymous referee of the  ``Foundation of Physics" which,
nevertheless, was of crucial importance in initiating my present
research. Many thanks to all them.

Zacatecas University, M\'exico, is thanked for awarding
the full professorship.  This work has been partly supported by
the Mexican Sistema Nacional de Investigadores, the Programa de Apoyo a la
Carrera Docente and by the CONACyT, M\'exico under the research project
0270P-E.

\end{document}